\documentclass[10pt,prd,aps,twocolumn,showpacs,showkeys
,superscriptaddress,groupedaddress,floatfix]{revtex4-1}
\usepackage{bm}
\usepackage{amsmath}
\usepackage{amsfonts}
\usepackage{amssymb}
\usepackage{graphicx}
\usepackage{graphics}
\usepackage[normalem]{ulem}
\usepackage{soul}
\usepackage[dvips]{color}
\usepackage[latin1]{inputenc}
\usepackage[english]{babel}
\usepackage{anysize}

\newcommand{\be}{\begin{equation}}
\newcommand{\ee}{\end{equation}}
\newcommand{\ba}{\begin{eqnarray}}
\newcommand{\ea}{\end{eqnarray}}

\def\ee{\mbox{$\left(e,e^{\prime}\right)$\ }}
\def\eep{\mbox{$\left(e,e^{\prime}p\right)$\ }}

\begin{document}

\author{Andrea Meucci}
\author{Carlotta Giusti}
\affiliation{Dipartimento di Fisica,
Universit\`{a} degli Studi di Pavia and \\
INFN,
Sezione di Pavia, via A. Bassi 6, I-27100 Pavia, Italy}

\title{Final-state interactions effects
in neutral-current neutrino and antineutrino cross sections at MiniBooNE kinematics}

\date{\today}

\begin{abstract}
The predictions of different relativistic descriptions of final-state interactions
are compared with the 
neutral current elastic neutrino and 
antineutrino-nucleus differential cross sections recently measured by the 
MiniBooNE Collaboration. 
The results of the relativistic Green's function model describe of the data 
without the need to increase the standard value of the 
nucleon axial mass.

\end{abstract}


\maketitle


\section{Introduction}
\label{intro}

The MiniBooNE Collaboration at FermiLab has reported 
measurements of the neutral-current elastic (NCE)
flux-averaged differential neutrino \cite{miniboonenc} and, more recently, 
antineutrino \cite{miniboonenc-nubar} cross
sections on CH$_2$ as a function of the four-momentum transferred squared 
$Q^2$.
Measurements of the neutrino and antineutrino charged-current quasielastic 
(CCQE) cross sections on carbon have also been reported in \cite{miniboone,miniboone-ant}.  

At the few GeV energy region considered in the MiniBooNE experiments 
theoretical models based on  the impulse approximation (IA)
are usually unable to reproduce the experimental cross 
sections \cite{Amaro:2006pr,Antonov:2007vd,Benhar:2011wy,Ankowski:2012ei}
unless calculations are performed with a
value of the nucleon axial mass $M_A$ significantly larger
($M_A \sim 1.20 \div 1.40$ GeV/$c^2$)  than the world average value from the
deuterium data of $M_A \simeq 1.03$ GeV/$c^2$~\cite{Bernard:2001rs,bodek08}.
This has been viewed as an indication that the reaction can have significant
contributions from effects beyond  the IA.
A careful analysis of all nuclear effects
and of the relevance of multinucleon emission and of some non-nucleonic
contributions is  required for a deeper understanding of the reaction 
dynamics \cite{PhysRevC.79.034601,Leitner:2010kp,
PhysRevC.83.054616,FernandezMartinez2011477,AmaroAntSusa,Amaro:2011qb,
Morfin:2012kn,Martini:2013sha,Nieves201390,Golan:2013jtj}.

Models developed for quasielastic (QE) electron 
scattering~\cite{Boffi:1993gs,book} and able to
successfully describe electron scattering data can  provide 
a useful tool to study neutrino-induced reactions. 
A reliable description of the effects of final-state interactions (FSI) between 
the ejected nucleon and the residual nucleus is very important for the 
comparison with data~\cite{Boffi:1993gs,book,Udias:1993xy,Meucci:2001qc,
Meucci:2001ja,Meucci:2001ty,Radici:2003zz,Giusti:2011it}.  
The relevance of FSI has 
been cleary stated in the case of the exclusive \eep  reaction within the 
distorted-wave impulse approximation (DWIA), where the use of a complex optical 
potential (OP) whose imaginary part produces an absorption that reduces the 
calculated cross section is essential to reproduce the data. 
In the case of the inclusive $\left(e,e^{\prime}\right)$ reaction, as well as of
CCQE and NCE neutrino scattering, the use of the DWIA with an absorptive 
complex OP is not correct and different approaches have been adopted. In the 
relativistic plane-wave impulse approximation (RPWIA) FSI are neglected.
In other approaches based on the IA FSI are incorporated in the final nucleon state 
either retaining only the real part of the relativistic optical potential (rROP), or 
using the same relativistic mean field
potential considered in describing the initial nucleon state (RMF) 
\cite{Maieron:2003df,Caballero:2005sn}.

In the relativistic Green's function (RGF) model the components of the nuclear response are written
in terms of matrix elements of the same type as the DWIA ones
of the exclusive \eep process, but involve
eigenfunctions of the OP and of its Hermitian conjugate, where the imaginary
part has an opposite sign and gives in one case an absorption and in the
other case a gain of strength. A detailed description of the model can be found
in our previous work \cite{Capuzzi:1991qd,Meucci:2003uy,Meucci:2003cv,Capuzzi:2004au,
Meucci:2005pk,Meucci:2009nm,Meucci:2011pi}. 


In \cite{Meucci:2011nc,PhysRevC.88.025502} we have already discussed  
the results of different relativistic descriptions
of FSI for NCE scattering averaged over the $\nu$ and $\bar{\nu}$ 
MiniBooNE flux.
We note that the RGF is appropriate for an inclusive process like CCQE,  
whereas the NCE scattering is a semi-inclusive 
process where the final neutrino cannot be detected and only the final nucleon  
is measured. As a consequence, the RGF may include channels which 
are not present in the experimental NCE measurements, but it can also recover 
important contributions which are not taken into account by other models based 
on the IA.
In comparison with the MiniBooNE NCE neutrino scattering data, the DWIA and 
RMF generally underpredict the experimental 
cross section, while the RGF results are in reasonable agreement with the 
data.
In this report we extend the comparison to MiniBooNE NCE antineutrino data 
which have only recently become available.



 \section{Results at MiniBooNE kinematics} \label{minib}

In all the calculations we have adopted the standard value 
of the axial mass, i.e., $M_A = 1.03$ GeV$/c^2$, and we have neglected  
strangeness effects in the nucleon form factors.
The bound nucleon states 
are taken as self-consistent Dirac-Hartree solutions derived 
within a relativistic mean field approach using a Lagrangian containing 
$\sigma$, $\omega$, and $\rho$ mesons 
\cite{Serot:1984ey,Rein:1989,Rign:1996,Lalazissis:1996rd,Serot:1997xg}.

In the RGF calculations we have used three parametrizations for the 
relativistic OP of $^{12}$C: the Energy-Dependent and A-Dependent EDAD1
(the $E$ represents the energy and the $A$ the atomic number) and the
Energy-Dependent but A-Independent EDAI OPs of \cite{Cooper:1993nx}, 
and  the more recent Democratic (DEM)
OP of \cite{Cooper:2009}. EDAI is a single-nucleus parametrization, which is 
constructed to better reproduce the elastic proton-$^{12}$C phenomenology. 
In contrast, EDAD1
and DEM depend on the atomic number $A$ and are obtained through 
a fit to more than 200 data sets of elastic proton-nucleus scattering data on a wide
range of nuclei that is not limited to doubly closed shell nuclei.
In \cite{esotici2,Meucci:2013pca} the evolution of 
QE electron scattering cross sections along different
isotopic and isotonic chains evaluated within the RGF with the DEM OP is 
presented: the different number of nucleons along the  chains
produces different OPs, but the results for the cross sections are always 
reasonable, even for nuclei with large asymmetry between the number of neutrons and protons.

\begin{figure}[t]
    \centering
        \includegraphics[scale=.42]{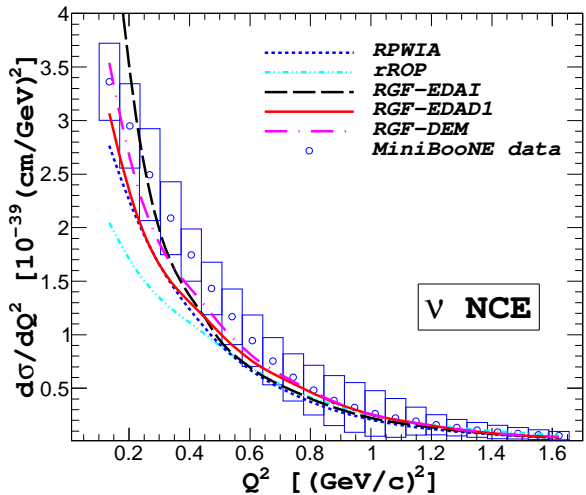}\vskip 3mm%
     \caption{(Color online)
    NCE flux-averaged $(\nu N \rightarrow \nu N)$  cross section as
    a function of $Q^2$. 
    The data are from \cite{miniboonenc}.}
    \label{fig:miniboone_models}
\end{figure}


In Fig.~\ref{fig:miniboone_models} our   
NCE $(\nu N \rightarrow \nu N)$ cross sections are displayed as a function of $Q^2=2m_N \sum_i T_i$,
where  $Q^2$ is defined assuming the
target nucleon  at rest, $m_N$ is the nucleon mass, and $\sum_i T_i$ is the 
total kinetic energy of the outgoing nucleons. 
These results have already been published 
in  \cite{Meucci:2011nc,PhysRevC.88.025502} and are presented here again for 
completeness. The RPWIA 
results, where FSI are neglected, show a satisfactory, although not perfect,
agreement with the magnitude of the data whereas
the rROP ones, which are  calculated retaining 
only the real part of the EDAI potential,  underestimate the 
data but for $Q^2 \geq 0.6$ (GeV$/c)^2$. 
The RGF cross sections are in better agreement with the data.
The differences between the RGF results 
are due to the different imaginary parts of the relativistic OPs adopted in the
calculations, which produce large differences in the neutrino-nucleus cross
sections evaluated at fixed neutrino energies and at small values 
of the outgoing nucleon kinetic energy,

\begin{figure}[t]
    \centering
     \includegraphics[scale=.42]{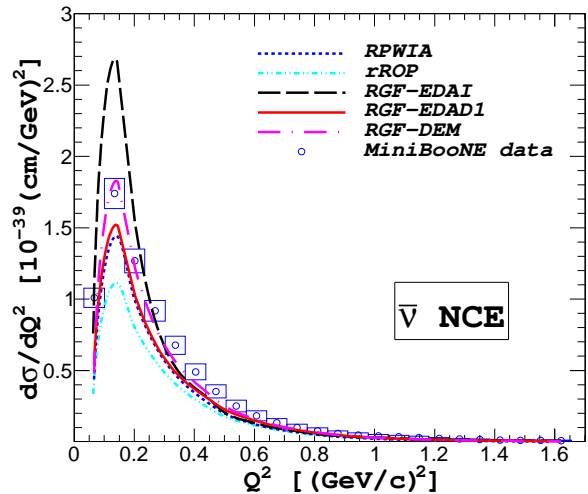} \vskip 3mm
    \caption{(Color online)
    The same as in Fig. \ref{fig:miniboone_models}, but for
    $(\bar{\nu} N \rightarrow \bar{\nu} N)$ cross section. 
    The data are from \cite{miniboonenc-nubar}.}
    \label{fig:miniboone_models_nubar}
\end{figure}


In Fig.~\ref{fig:miniboone_models_nubar} we present  our flux-averaged NCE 
$(\bar{\nu} N \rightarrow \bar{\nu} N)$ cross sections as a function 
of $Q^2$.
In \cite{PhysRevC.88.025502} we already presented theoretical predictions 
for antineutrino cross sections using the set of efficiency coefficients
given in \cite{miniboonenc} for neutrino scattering. The 
selection for the antineutrino NCE sample is slightly different from the neutrino
one and the efficiencies are only approximately similar. The results displayed
in Fig.~\ref{fig:miniboone_models_nubar} are evaluated with the correct 
efficiencies for antineutrinos.
Also in the antineutrino case the rROP cross sections
are lower than the RPWIA ones, whereas the RGF  produces
larger cross sections. The data are underpredicted by the rROP  
and satisfactorily described by the RPWIA.
A better agreement with the data is generally provided by the RGF, 
in particular when the DEM OP is adopted. The RGF-EDAI results are  
enhanced at $Q^2 \approx 0.1$ (GeV$/c)^2$ and the RGF-EDAD1 cross sections are 
similar to the RPWIA ones 
in the entire kinematical range of MiniBooNE $\bar{\nu}$ flux.
All the models are able to reasonably reproduce the first datum at 
$Q^2 \approx 0.06$ (GeV$/c)^2$.

\begin{figure}[t]
    \centering
     \includegraphics[scale=.42]{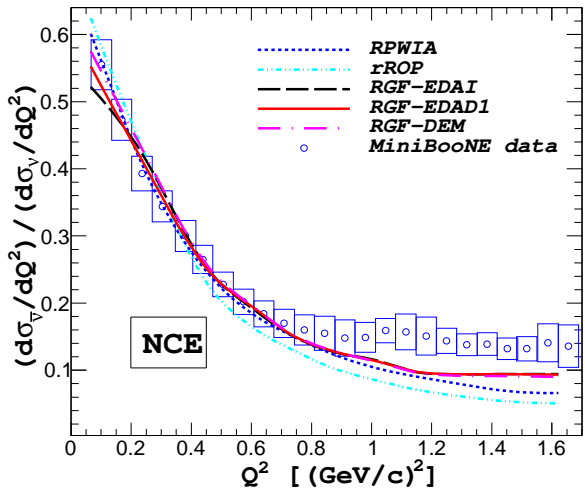}\vskip 3mm
    \caption{(Color online)
    Ratio of the $\bar{\nu}$ to $\nu$
NCE scattering cross section with total error.  
 The data are from \cite{miniboonenc-nubar}.}
    \label{fig:ratio}
\end{figure}


In Fig.~\ref{fig:ratio} we show our results
for the ratio of the $\bar{\nu}$ to $\nu$ NCE scattering cross section.
Ratios of cross sections are particularly interesting, owing to the fact that 
in the ratios distortion effects are largely reduced and different descriptions
of FSI are expected to produce similar 
results \cite{Meucci:2004ip,Meucci:2006ir,PhysRevC.88.025502}.
From the experimental point of view, both $\nu$ and $\bar{\nu}$
NCE measurements were made in the same beamline and with the same detector but
with opposite horn polarities \cite{miniboonenc,minibooneflux,miniboonenc-nubar}.
Despite the fact that the experimental $\nu$ and $\bar{\nu}$ fluxes are not 
identical,  the ratio of the two cross 
sections should minimize the errors and  provide a useful observable to 
test various theoretical models. 

In Fig.~\ref{fig:ratio} all the models give, as it was expected, very close 
results. In particular, the RGF results are practically independent of the
parametrization adopted for the OP; 
small differences can be seen only at very low $Q^2$, but all the results are 
within the experimental errors.
At large $Q^2$ all the results slightly underestimate the data. This is related to 
the fact that the $\bar{\nu}$ cross sections in 
Fig. \ref{fig:miniboone_models_nubar} are underestimated for  
$Q^2 \geq 1$ (GeV$/c)^2$, whereas the $\nu$ cross sections in 
Fig. \ref{fig:miniboone_models} are within the error bars in the entire range 
of $Q^2$.\\

\section{Conclusions}

The predictions of
different relativistic descriptions of FSI have been compared with the NCE 
MiniBooNE data for $\nu$ and $\bar{\nu}$-nucleus scattering . 

The RGF results are able to describe the data, both for $\nu$ and $\bar{\nu}$
scattering, without the need to increase the standard value of the
axial mass. Visible differences on the RGF cross sections at low 
values of $Q^2$  are produced by the use of different parametrizations 
of the phenomenological OP. The best agreement with the data is given by the
RGF-DEM results.  In the ratio of the  $\bar{\nu}$ to $\nu$ NCE scattering 
cross sections FSI effects are strongly reduced and 
the RGF results are practically independent of the choice of the OP. 
The experimental ratio is reasonably described 
when $Q^2 \leq 1$ (GeV$/c)^2$  and slightly underestimated when   
$Q^2 \ge 1$ (GeV$/c)^2$ by all the RGF results. 

Different and independent models have confirmed the important role of 
contributions other than direct one-nucleon emission.
Owing to the flux-average procedure, $\nu$-nucleus reactions at 
MiniBooNE can have important contributions from effects beyond the IA in some 
kinematic regions where the experimental $\nu$ flux has significant 
strength \cite{Benhar:2010nx,Benhar:2011wy}.
Although contributions like, e.g., rescattering processes of the nucleon, 
non-nucleonic $\Delta$ excitations, or  multinucleon processes  are not 
included explicitly in the RGF model, they can be recovered, at least to some 
extent, by the imaginary part of the phenomenological OP. The use of such a 
phenomenological ingredient, however, does not allow us to explain in detail 
the origin of the enhancement of the RGF results with respect to the results 
of other models based on the IA. 
At present, lacking a phenomenological OP which exactly fulfills the 
dispersion relations in the whole energy region of interest, the RGF predictions are
not univocally determined from the elastic phenomenology.  
A better determination of a relativistic OP, which closely fulfills the
dispersion relations, deserves further investigation.

%

\end{document}